\documentstyle[11pt]{article}
\begin{document} 
\makeatletter
\@addtoreset{equation}{section}
\makeatother
\renewcommand{\theequation}{\thesection.\arabic{equation}}
\baselineskip 15pt

\title{\bf Which Kind of Two-Particle States Can Be Teleported through a
Three-Particle Quantum Channel?
\footnote{Work supported in part by Istituto Nazionale di Fisica Nucleare,
 Sezione di Trieste, Italy}}
\author{Luca Marinatto\footnote{e-mail: marinatto@ts.infn.it}\\
{\small Department of Theoretical Physics of the University of Trieste, and}\\
{\small Istituto Nazionale di Fisica Nucleare, Sezione di Trieste, Italy.}\\
and \\
\\ Tullio Weber\footnote{e-mail: weber@ts.infn.it}\\
{\small Department of Theoretical Physics of the University of Trieste, and}\\
{\small Istituto Nazionale di Fisica Nucleare, Sezione di Trieste, Italy.}}

\date{}

\maketitle

\begin{abstract}
The use of a three-particle quantum channel to teleport entangled states
 through a slight modification of the standard teleportation procedure is
 studied. It is shown that it is not possible to perform successful
teleportation of an arbitrary and unknown two-particle entangled state,
 following our version of the standard teleportation procedure. On the 
contrary, it is shown which, and in how many different ways, particular
 classes of two-particle states can be teleported. \\

Key words: Teleportation, Bell measurement, Entanglement.

\end{abstract}

%---------------------------------------------------------------------------

\section{Introduction.}

The quantum teleportation process permits one to transmit unknown 
 quantum states from a sender to a receiver which are spatially
 separated.

The classical idea of teleportation, as portrayed in science fiction novels and
movies, involves a complete dematerialization of an object positioned in place
 A and its reappearence at a distant place B.

By contrast, quantum teleportation differs from this fanciful idea since
 it is only possible to teleport from A to B the state representing one or
 more particles in A, by transferring it to one or more particles already
 existing in B.

One possible way of performing such a process consists in firstly learning
 all the properties of the original and unknown state, and then in transmitting
 them, by means of classical communication channels, to a receiver able to
recreate a perfect copy.

Nevertheless such a procedure would not work, for it would be necessary
 to have an
 infinite ensemble of identically prepared quantum systems in order to
 completely determine their original state.

However one of the most striking features of Quantum Mechanics, i.e. the
 entanglement, supplies us with a quantum channel of communication being
able to transfer an unknown quantum state
from a place to another, without knowing it and without violating special
relativity constraints (i.e., not instantaneously).

Following the original proposal of Bennett et al. \cite{ref1}, to teleport
a single-particle state the sender,
traditionally named Alice, and the receiver, Bob, need only to share a
particular two-particle maximally entangled state (an EPR singlet state)
 which acts as a purely quantum channel, and a telephone, the classical
 channel.

The core of the teleportation process resides on a projective local
 measurement, performed by Alice, in the Bell basis consisting of four
 orthonormal and maximally entangled two-particle states, involving the
unknown one.
After sending the result of the measurement by the classical channel ( the
crucial step which prevents from sending faster-than light messages )
Bob can reconstruct, applying the unitary and local transformations 
suggested by Alice, a perfect copy of the original state.
At the end the unknown state of Alice is destroyed, respecting in a such way
the no-cloning theorem \cite{ref2}.

Bennett et al. applied this type of teleportation to single-particle states
 suggesting however that it should have been working also for arbitrary
 $N$-particle entangled states. As a matter of fact, it works through the use
 of $N$ two-particle quantum channels. 

 In the present work we explore the possibility of obtaining the same result 
using only
one quantum channel, realized by peculiar kinds of three-particle
states, shared by Alice and Bob: the sender possesses one of the particles
 while the remaining two, belonging to the receiver, will be used as building
 blocks for the copy of the teleported state. We note that three-particle
 states have been obtained experimentally and could be used in the present
 context \cite{ref3}.

As it will appear, only some kinds of two-particle entangled states, and
not the most general one, can be teleported by means of three-particle quantum 
channels and measurements onto two-particle Bell entangled states.

In appendix A all the different ways to achieve permitted teleportation
are enumerated.

\section{Teleportation of an arbitrary state}

Let us suppose that Alice wants to teleport to Bob, spacelike separated from
her, an unknown and arbitrary two-particle entangled state$\,$:

\begin{equation}
\label{stato}
\left\{ \begin{array}{ll}
        \vert\psi\rangle_{12}=\alpha\vert 00 \rangle_{12} + \beta\vert 10
\rangle_{12} + \delta\vert 01 \rangle_{12} + \gamma\vert 11 \rangle_{12}  \\
        \vert\alpha\vert^{2} + \vert\beta\vert^{2} + \vert\delta\vert^{2} + 
        \vert\gamma\vert^{2} = 1
        \end{array}
\right.
\end{equation}

We are following the Quantum Information Theory convention of indicating with 
$ \{ \vert 0 \rangle , \vert 1 \rangle \} $ a complete orthonormal basis for a
two dimensional, single particle, Hilbert space, made up of eigenstates of
operator $ \sigma_{z} $ with eigenvalues respectively $ +1 $ and $ -1 $.
In order to teleport a two-particle state, the simplest single quantum-channel 
which can be used must be a three-particle one$\,$:

\begin{equation}
 \vert\phi\rangle_{345} = \frac{1}{\sqrt{N}}\, [\, a\vert 000 \rangle +
  b\vert 100 \rangle + c\vert 010 \rangle + d\vert 001 \rangle +
  e\vert 110 \rangle + f\vert 101 \rangle + g\vert 011 \rangle +
  h\vert 111 \rangle \,]_{345}
\end{equation}

with $ a,b...h = \pm 1$ or $0$, and $N$ equal to the number of coefficients
which are not zeros.

 The initial state we start from, will be the product of the two states just
 defined$\,$:

\begin{equation}
\label{statoiniziale}
 \vert \Omega \rangle = \vert\psi\rangle_{12}\: \vert\phi\rangle_{345}
\end{equation} 
%\vspace{0.2cm}
 
where particles labelled $1,2$ and $3$ belong to Alice ( she can perform
local measurements only on them ), while particles labelled $4$ and $5$
are placed near Bob and they will constitute the bricks to make
 a copy of the state $\vert \psi \rangle$.

The core of the teleportation mechanism resides in performing all the direct
products between states in $\vert\psi\rangle_{12}$ and $\vert\phi\rangle_{345}$
and in expressing subsequently the Alice's states of particles $2$ and $3$ in
terms of vectors of the Bell basis.

This complete basis consists of four orthonormal maximally entangled states,
simultaneous eigenvectors of the two commuting operators,
$(\sigma_{x})_{2}(\sigma_{x})_{3}$ and $(\sigma_{z})_{2}(\sigma_{z})_{3}\,$:

\begin{equation}
\label{bellbasis}
 \vert \phi^{+} \rangle_{23}= \frac{1}{\sqrt{2}} \left( \vert 00 \rangle + 
\vert 11 \rangle \right)_{23} \:\:\:\:\:\:\:\:\:  \vert \psi^{+} \rangle_{23}= 
\frac{1}{\sqrt{2}} \left( \vert 01 \rangle + \vert 10 \rangle \right)_{23}
\end{equation}
\[
\vert \phi^{-} \rangle_{23}= \frac{1}{\sqrt{2}} \left( \vert 00 \rangle - 
\vert 11 \rangle \right)_{23} \:\:\:\:\:\:\:\:\:  \vert \psi^{-} \rangle_{23}= 
\frac{1}{\sqrt{2}} \left( \vert 01 \rangle - \vert 10 \rangle \right)_{23} \]

Inserting expressions (\ref{bellbasis}) into equation (\ref{statoiniziale}) the
 resulting state of the system -- completely equivalent to the original one
 since no physical process has taken place -- is$\,$:

\begin{eqnarray}
\label{senzadamo}
    \vert \Omega \rangle & = & \frac{1}{\sqrt{2N}} \vert \phi^{+} \rangle_{23}
 \left\{  \vert 0 \rangle_{1} \left[ (\alpha a + \delta b) \vert 00 \rangle +
 (\alpha c + \delta e) \vert 10 \rangle + (\alpha d + \delta f) \vert 01 
 \rangle + (\alpha g + \delta h) \vert 11 \rangle \right]_{45} + \right.
  \nonumber  \\
 & & \:\:\:\:\:\:\:\:\:\:\:\:\:\:\:\:\:\:\:\:\: + \left. \vert 1 \rangle_{1}
 \left[
 (\beta a +\gamma b) \vert 00 \rangle +
 (\beta c +\gamma e) \vert 10 \rangle + (\beta d + \gamma f) \vert 01 \rangle +
 (\beta g + \gamma h) \vert 11 \rangle \right]_{45} \right\} +  \nonumber \\
 & &   \frac{1}{\sqrt{2N}} \vert \phi^{-} \rangle_{23} \left\{ \vert 0
 \rangle_{1} \left[ (\alpha a -\delta b) \vert 00 \rangle + (\alpha c - \delta
 e) \vert 10 \rangle + (\alpha d - \delta f) \vert 01 \rangle + (\alpha g -
 \delta h) \vert 11\rangle \right]_{45} +  \right.  \nonumber \\
 & & \:\:\:\:\:\:\:\:\:\:\:\:\:\:\:\:\:\:\:\:\: \left. + \vert 1\rangle_{1} 
\left[ (\beta a -\gamma b) \vert 00 \rangle +
 (\beta c - \gamma e) \vert 10 \rangle + (\beta d - \gamma f) \vert 01 \rangle
 + (\beta g -\gamma h) \vert 11 \rangle \right]_{45} \right\} + \nonumber \\ 
 & &   \frac{1}{\sqrt{2N}} \vert \psi^{+} \rangle_{23} \left\{ \vert 0 
\rangle_{1} \left[ (\alpha b +\delta a) \vert 00 \rangle + (\alpha e + \delta c)
 \vert 10 \rangle + (\alpha f + \delta d) \vert 01 \rangle + (\alpha h + \delta 
g) \vert 11 \rangle \right]_{45} + \right. \nonumber \\
 & & \:\:\:\:\:\:\:\:\:\:\:\:\:\:\:\:\:\:\:\:\: \left. + \vert 1 \rangle_{1} 
\left[ (\beta b +\gamma a) \vert 00 \rangle +
 (\beta e + \gamma c) \vert 10 \rangle + (\beta f + \gamma d) \vert 01 \rangle
  + (\beta h +\gamma g) \vert 11\rangle \right]_{45} \right\} +\nonumber \\
 & &   \frac{1}{\sqrt{2N}} \vert \psi^{-} \rangle_{23} \left\{ \vert 0 
\rangle_{1}  \left[ (\alpha b -\delta a) \vert 00 \rangle + (\alpha e - \delta 
c) \vert 10\rangle + (\alpha f - \delta d) \vert 01 \rangle + (\alpha h - \delta 
g) 
\vert 11 \rangle \right]_{45}  \right. + \nonumber \\
 & & \:\:\:\:\:\:\:\:\:\:\:\:\:\:\:\:\:\:\:\:\: \left. + \vert 1 \rangle_{1}
  \left[ (\beta b -\gamma a) \vert 00 \rangle +
 (\beta e - \gamma c) \vert 10 \rangle + (\beta f -\gamma d) \vert 01 \rangle + 
(\beta h - \gamma g) \vert 11 \rangle \right]_{45} \right\} \nonumber \\
 & &  
\end{eqnarray}

Alice's strategy to pursue the teleportation process, as already said,
 will consist in a local projective measurement onto the vectors of the Bell
 basis of particles $2$ and $3$ and in a successive measurement on particle
 $1$, to be specified, in order to leave Bob's particles $4$ and $5$ in a state
looking very much like Alice's original one.

As we are going to show, it turns out that there is no possible choice of the
coefficients $a,b....h$, up to now deliberately unspecified, fulfilling
the desired task of complete teleportation : it will not be possible for Bob
 to reconstruct the state $\vert \psi\rangle$.

First of all it is worthwhile observing that in equation (\ref{senzadamo})
 the couples of coefficients $(\alpha,\delta)$ and $(\beta, \gamma)$ are only
 and always associated to states $\vert 0\rangle_{1}$ and
 $\vert 1\rangle_{1}$, respectively.

It is so clear that successive measurement on Bell basis for particles $2$
 and $3$ and on canonical basis of particle $1$, leave the collapsed state
 quite dissimilar from the one to teleport, because of the lack of needed
coefficients.

Suppose for example Alice obtains the state $\vert\phi^{+}\rangle_{23}$ and the
state $\vert 0\rangle_{1}$ in successive measurements: in the remaining state
of the system there will be no trace of $\beta$ and $\gamma$, preventing
 Bob from reconstructing state $\vert\psi\rangle$.
In order to overcome this difficulty, Alice must resort to perform an Hadamard
transformation on particle $1$, so mixing up the coefficients.

We note that Alice could obtain the same result by making the substitution$\,$:

\begin{equation}
\vert 0 \rangle_{1} = A \vert \varphi \rangle_{1} + B \vert \chi \rangle_{1} 
\end{equation}
\[
 \vert 1 \rangle_{1} = -B^{\star} \vert \varphi \rangle_{1} + A^{\star} \vert
 \chi \rangle_{1} 
\]

In fact, in order not to have undesired coefficients in the transmitted state,
 she should be obliged to choose $A$=$B$=$1/ \sqrt{2}$, obtaining an
 expression having exactly the same distinctive features of equation
 (\ref{conadamo}).
This means that making the Hadamard transformation or projecting on states
 $\vert \varphi_{1}\rangle_{1} $ and $\vert \chi_{1}\rangle_{1}$ has the same
 effect.

Hadamard unitary transformation is defined in terms of Pauli matrices as$\,$:

\begin{equation}
 H =\frac{1}{\sqrt{2}}\, (\, \sigma_{x} + \sigma_{z}\,) \:\:\:\:\:\:\:\:\:\:\:
H^{2}= 1
\end{equation}

It acts as a rotation about the axis $\vec{n} =1/\sqrt{2} \,( \vec{n_{x}} +
\vec{n_{z}})\,$:

\begin{equation}
 \left\{ \begin{array}{ll}
  H \,\vert 0\rangle = \frac{1}{\sqrt{2}}\,(\, \vert 0\rangle + \vert 1\rangle
 \,) \\
  H \,\vert 1\rangle = \frac{1}{\sqrt{2}}\,(\, \vert 0\rangle - \vert 1\rangle
 \,)
 \end{array}
 \right. 
\end{equation}

After applying such a transformation, the resulting state of the system with
mixed coefficients, is$\,$:

\begin{eqnarray}
\label{conadamo}
 \vert \tilde{\Omega} \rangle & = & \frac{1}{2\sqrt{N}} \vert \phi^{+} 
\rangle_{23}
 \left\{ \vert 0 \rangle_{1} \left[ \left( (\alpha +\beta)a + (\delta +
 \gamma)b \right) \vert 00 \rangle + \left( (\alpha +\beta)c + (\delta +
 \gamma)e \right) \vert 10 \rangle + \right. \right. \nonumber\\
&&\:\:\:\:\:\:\:\:\:\:\:\:\:\:\:\:\:\:\:\:\:\:\:\:\:\:\:\:\:\:\:\:\left. +
 \left( (\alpha +\beta)d + (\delta + \gamma)f \right) \vert 01 \rangle +
 \left( (\alpha + \beta)g + (\delta + \gamma)h \right) \vert 11 \rangle
 \right]_{45} +\nonumber \\
&&\:\:\:\:\:\:\:\:\:\:\:\:\:\:\:\:\:\:\:\:\:\:\:\:\:\:\:\: \vert 1 \rangle_{1} 
\left[ \left( (\alpha - \beta)a + (\delta -\gamma)b 
 \right) \vert 00 \rangle + \left( (\alpha -\beta)c + (\delta -\gamma)e \right)
 \vert 10 \rangle \right. + \nonumber\\
&&\:\:\:\:\:\: \:\:\:\:\:\:\:\:\:\:\:\:\:\:\:\:\:\:\:\:\:\:\:\:\:\:\left. \left. 
+
 \left( (\alpha -\beta)d + (\delta - \gamma)f \right)
 \vert 01\rangle + \left( (\alpha -\beta)g + (\delta -\gamma)h \right)
 \vert 11\rangle \right]_{45} \right\} +\nonumber \\
&& \frac{1}{2\sqrt{N}} \vert \phi^{-} \rangle_{23} \left\{ \vert 0 \rangle_{1}
 \left[ \left( (\alpha +\beta)a - (\delta +\gamma)b \right) \vert 00 \rangle
 + \left( (\alpha +\beta)c - (\delta + \gamma)e \right) \vert 10 \rangle + 
\right. \right.\nonumber \\
&&\:\:\:\:\:\:\:\:\:\:\:\:\:\:\:\:\:\:\:\:\:\:\:\:\:\:\:\:\:\:\:\: \left. +
 \left( (\alpha +\beta)d - (\delta + \gamma)f \right)
 \vert 01\rangle + \left( (\alpha + \beta)g - (\delta +\gamma)h \right)
 \vert 11\rangle \right]_{45} + \nonumber\\
&& \:\:\:\:\:\:\:\:\:\:\:\:\:\:\:\:\:\:\:\:\:\:\:\:\:\:\:\:\vert 1 \rangle_{1} 
\left[ \left( (\alpha -\beta)a - (\delta -\gamma)b 
 \right) \vert 00 \rangle + \left( (\alpha -\beta)c - (\delta -\gamma)e \right)
 \vert 10\rangle + \right.\nonumber \\
&& \:\:\:\:\:\:\:\:\:\:\:\:\:\:\:\:\:\:\:\:\:\:\:\:\:\:\:\:\:\:\:\:\left.
 \left.+
 \left( (\alpha -\beta)d - (\delta -\gamma)f \right)
 \vert 01\rangle + \left( (\alpha -\beta)g - (\delta -\gamma)h \right)
 \vert 11\rangle \right]_{45} \right\} +\nonumber \\
&& \frac{1}{2\sqrt{N}} \vert \psi^{+} \rangle_{23} \left\{ \vert 0\rangle_{1}
 \left[ \left( (\alpha +\beta)b + (\delta +\gamma)a 
 \right)\vert 00\rangle + \left( (\alpha +\beta)e + (\delta + \gamma)c \right)
 \vert 10 \rangle + \right. \right. \nonumber\\
&&\:\:\:\:\:\:\:\:\:\:\:\:\:\:\:\:\:\:\:\:\:\:\:\:\:\:\:\:\:\:\:\:\left. +
 \left( (\alpha +\beta)f + (\delta + \gamma)d \right)
 \vert 01 \rangle + \left( (\alpha + \beta)h + (\delta +\gamma)g \right)
 \vert 11 \rangle \right]_{45} +\nonumber\\  
&&  \:\:\:\:\:\:\:\:\:\:\:\:\:\:\:\:\:\:\:\:\:\:\:\:\:\:\:\: \vert 1 \rangle_{1} 
\left[ \left( (\alpha -\beta)b + (\delta -\gamma)a 
 \right)\vert 00\rangle + \left( (\alpha -\beta)e + (\delta -\gamma)c \right)
 \vert 10 \rangle + \right.\nonumber \\
&&\:\:\:\:\:\:\:\:\:\:\:\:\:\:\:\:\:\:\:\:\:\:\:\:\:\:\:\:\:\:\:\:
 \left. \left. +
 \left( (\alpha -\beta)f + (\delta -\gamma)d \right)
 \vert 01 \rangle + \left( (\alpha -\beta)h + (\delta -\gamma)g \right)
 \vert 11 \rangle \right]_{45} \right\} +\nonumber \\
&&
 \frac{1}{2\sqrt{N}} \vert \psi^{-} \rangle_{23} \left\{
 \vert 0 \rangle_{1}  \left[  \left( (\alpha +\beta)b - (\delta +\gamma)a 
 \right) \vert 00\rangle + \left( (\alpha +\beta)e - (\delta + \gamma)c \right)
 \vert 10 \rangle + \right. \right. \nonumber\\
&&\:\:\:\:\:\:\:\:\:\:\:\:\:\:\:\:\:\:\:\:\:\:\:\:\:\:\:\:\:\:\:\: \left. +
 \left( (\alpha +\beta)f - (\delta + \gamma)d \right)
 \vert 01 \rangle + \left( (\alpha + \beta)h - (\delta +\gamma)g \right)
 \vert 11 \rangle \right]_{45} + \nonumber\\
&& \:\:\:\:\:\:\:\:\:\:\:\:\:\:\:\:\:\:\:\:\:\:\:\:\:\:\:\:\vert 1 \rangle_{1} 
\left[ \left( (\alpha -\beta)b - (\delta -\gamma)a 
 \right) \vert 00\rangle + \left( (\alpha -\beta)e - (\delta -\gamma)c \right)
 \vert 10 \rangle + \right.\nonumber \\
&&\:\:\:\:\:\:\:\:\:\:\:\:\:\:\:\:\:\:\:\:\:\:\:\:\:\:\:\:\:\:\:\:
 \left. \left. +  \left( (\alpha -\beta)f - (\delta -\gamma)d \right)
\vert 01 \rangle + \left( (\alpha -\beta)h - (\delta -\gamma)g \right)
 \vert 11 \rangle \right]_{45} \right\}\nonumber \\
&& 
\end{eqnarray}

Alice now accomplishes Bell measurement on particles $2$ and $3$, and canonical
measurement on particle $1$, projecting in a such way the state firstly 
on Bell basis $\{ \vert \phi^{\pm} \rangle\, , \vert \psi^{\pm} \rangle \}$
and secondly onto the canonical basis $\,\{ \vert 0\rangle,\vert 1\rangle \}$:
 the state of the system will eventually collapse, with equal probability,
 in one of the eight states of particles $4$ and $5$ indicated between square
 brackets. 

It is clear that there is no possible choice of $a, b... h$ suited to
 reproduce the unknown initial state $\vert \psi \rangle$, because the
 coefficients $\alpha,\beta, \delta, \gamma$ -- although mixed -- remains
 summed two by two ($\alpha\pm \beta$ and $\delta \pm \gamma$).

 Then, it is not possible to achieve complete teleportation of an arbitrary and
 unknown two-particle entangled state ( i.e. with $\alpha, \beta, \gamma, \delta 
\neq 0$) using the present scheme.

One can then ask whether a measurement on orthonormal states $\vert \phi_{i} 
\rangle_{23}$, with $i=1...4$, different from the Bell states, could make one to
 achieve the desired goal.
A cumbersome but trivial calculation shows that there is no gain on projecting
 onto general states and in all cases teleportation cannot be obtained.

In particular, since the proof is valid for every choice of non zero
 coefficients $\alpha, \beta, \gamma, \delta$, it is not even possible to
 teleport a two-particle factorized state using the present procedure.
In fact, each state of the type $(m\vert 0 \rangle_{1} + n\vert 1\rangle_{1})
(r\vert 0 \rangle_{2} + s \vert 1\rangle_{2})$ can be written in the form
 (\ref{stato}); conversely, it can be trivially shown that every state of
 the form (\ref{stato}), for which the condition $\alpha \delta = \gamma
 \beta $ is satisfied, can be expressed as a factorized state.
 
This result is not surprising. In fact, Alice has at her disposal only one
particle, i.e. one e-bit, which is not sufficient to teleport a general state
 of two particles. However, in all these cases, one can obtain successful
 teleportation by simply repeating the original standard teleportation
 procedure using a sequence of two (two-particle) channels, rather than a
 single (three-particle) quantum channel as considered in this article.

\section{Teleportation of peculiar states}

Since the present mechanism for teleportation cannot work for an
arbitrary two-particle entangled state, let us try to focus our attention
on some two dimensional subspaces of the whole Hilbert space of the system of
the two particles ( which is four dimensional ).
Let us try for example with the state $\alpha \vert 00 \rangle + \gamma \vert
11 \rangle$. 

Alice is now able to successfully perform the teleportation process by choosing
a suitable quantum channel and then following the steps already considered in
the previous section: 

\begin{enumerate}
\item Alice prepares the state $\vert \Omega \rangle = (\alpha \vert 00
\rangle_{12} + \gamma \vert 11 \rangle_{12}\,)\, \vert \phi \rangle_{345}$ where 
$\vert \phi \rangle_{345}$ is the three-particle quantum channel, yet to be 
specified. 
\item Alice acts with Hadamard unitary trasformation on states of particle $1$,
in order to mix up in an appropriate way the coefficients $\alpha$ and 
$\gamma$.
\item Alice performs a Bell measurement on particles $2$ and $3$.
\item Alice performs a measurement onto states $\vert 0 \rangle_{1}$ and $\vert
1 \rangle_{1}$. 
\end{enumerate}

The eight equally probable results are easily obtained by putting
 $\beta$=$\delta$=$0$ in equation (\ref{conadamo})$\,$:

\begin{itemize} 
\item  $ \vert 0 \rangle_{1} \vert \phi^{+} \rangle_{23} \:\: \Rightarrow \:\:
 (\alpha a + \gamma b) \vert 00 \rangle_{45} + (\alpha c + \gamma e) \vert 10
 \rangle_{45} + (\alpha d + \gamma f) \vert 01 \rangle_{45} + (\alpha g +
 \gamma h) \vert 11 \rangle_{45} $  
\item $
 \vert 1 \rangle_{1} \vert \phi^{+} \rangle_{23} \:\: \Rightarrow \:\:
 (\alpha a - \gamma b) \vert 00 \rangle_{45} + (\alpha c - \gamma e)\vert 10
 \rangle_{45} + (\alpha d - \gamma f) \vert 01 \rangle_{45} + (\alpha g -
 \gamma h) \vert 11 \rangle_{45} $ 
\item $
 \vert 0 \rangle_{1} \vert \phi^{-} \rangle_{23} \:\: \Rightarrow \:\:
 (\alpha a - \gamma b)\vert 00 \rangle_{45} + (\alpha c - \gamma e)\vert 10
 \rangle_{45} + (\alpha d - \gamma f) \vert 01 \rangle_{45} + (\alpha g -
 \gamma h) \vert 11 \rangle_{45} $ 
\item $
 \vert 1 \rangle_{1} \vert \phi^{-} \rangle_{23} \:\: \Rightarrow \:\:
 (\alpha a + \gamma b) \vert 00 \rangle_{45} + (\alpha c + \gamma e)\vert 10
 \rangle_{45} + (\alpha d + \gamma f) \vert 01 \rangle_{45} + (\alpha g +
 \gamma h) \vert 11 \rangle_{45} $ 
\item $
 \vert 0 \rangle_{1} \vert \psi^{+} \rangle_{23} \:\: \Rightarrow \:\:
 (\alpha b + \gamma a) \vert 00 \rangle_{45} + (\alpha e + \gamma c)\vert 10
 \rangle_{45} + (\alpha f + \gamma d) \vert 01 \rangle_{45} + (\alpha h +
 \gamma g) \vert 11 \rangle_{45} $
\item $
 \vert 1 \rangle_{1} \vert \psi^{+} \rangle_{23} \:\: \Rightarrow \:\:
 (\alpha b - \gamma a) \vert 00 \rangle_{45} + (\alpha e - \gamma c)\vert 10
 \rangle_{45} + (\alpha f - \gamma d) \vert 01 \rangle_{45} + (\alpha h -
 \gamma g) \vert 11 \rangle_{45} $
\item $
 \vert 0 \rangle_{1} \vert \psi^{-} \rangle_{23} \:\: \Rightarrow \:\:
 (\alpha b - \gamma a)\vert 00 \rangle_{45} + (\alpha e - \gamma c)\vert 10
 \rangle_{45} + (\alpha f - \gamma d) \vert 01 \rangle_{45} + (\alpha h -
 \gamma g) \vert 11 \rangle_{45} $
\item $
 \vert 1 \rangle_{1} \vert \psi^{-} \rangle_{23} \:\: \Rightarrow \:\:
 (\alpha b + \gamma a) \vert 00 \rangle_{45} + (\alpha e + \gamma c) \vert 10
 \rangle_{45} + (\alpha f + \gamma d) \vert 01 \rangle_{45} + (\alpha h +
 \gamma g) \vert 11 \rangle_{45}  $
\end{itemize}

There are now eight possible ways of choosing coefficients $a,b... h$ -- so
 there are eight quantum channels shared between Alice and Bob -- and,
 correspondingly, there are eight different sets of instructions to send via
 classical communications to Bob, in order to complete the teleportation
 process (see Appendix~A).

After having received that, Bob can successfully reconstruct the original state
by applying local unitary transformations on his particles.

Les us focus for example on the choice $a = h = 1\,\,\, , \,\,\,
 b=c=d=e=f=g=0$, corresponding to the quantum channel $\vert \phi \rangle =
1/\sqrt{2}\, (\vert 000 \rangle + \vert 111 \rangle )$.
We list below the set of instructions for Bob according to the results of
 Alice's measurements$\,$:

\vspace{0.5cm}
\begin{tabular}{c|c|c}
Alice's measurements     &      Bob's states    &     Bob's instructions \\
\hline
$\vert 0 \rangle_{1}\,\, \vert \phi^{+} \rangle_{23}\:\:\:$ & $\:\:\: \alpha 
\vert 00 \rangle_{45} + \gamma \vert 11 \rangle_{45}\:\:\:$ &   do nothing \\
$\vert 1 \rangle_{1}\,\, \vert \phi^{+} \rangle_{23}\:\:\:$ & $\:\:\:\alpha 
\vert 00 \rangle_{45} - \gamma \vert 11 \rangle_{45}\:\:\:$ &   apply 
$(\sigma_{z})_{4}\otimes I_{5}$ \\
$\vert 0 \rangle_{1}\,\, \vert \phi^{-} \rangle_{23}\:\:\:$ & $\:\:\:\alpha 
\vert 00 \rangle_{45} - \gamma \vert 11 \rangle_{45}\:\:\: $ &   apply 
$I_{4}\otimes (\sigma_{z})_{5}$ \\
$\vert 1 \rangle_{1}\,\, \vert \phi^{-} \rangle_{23}\:\:\:$ & $\:\:\:\alpha 
\vert 00 \rangle_{45} + \gamma \vert 11 \rangle_{45}\:\:\:$ &   do nothing \\
$\vert 0 \rangle_{1}\,\, \vert \psi^{+} \rangle_{23}\:\:\:$ & $\:\:\:\gamma 
\vert 00 \rangle_{45} + \alpha \vert 11 \rangle_{45\:\:\:}$ &   apply 
$(\sigma_{x})_{4}\otimes (\sigma_{x})_{5}$ \\
$\vert 1 \rangle_{1}\,\, \vert \psi^{+} \rangle_{23}\:\:\:$ & $\:\:\:-\gamma 
\vert 00\rangle_{45} + \alpha \vert 11 \rangle_{45}\:\:\:$ &   apply
$(\sigma_{z} \sigma_{x})_{4} \otimes (\sigma_{x})_{5}$ \\
$\vert 0 \rangle_{1}\,\, \vert \psi^{-} \rangle_{23}\:\:\:$ & $\:\:\:-\gamma 
\vert 00\rangle_{45} + \alpha \vert 11 \rangle_{45}\:\:\:$ &   apply
$(\sigma_{x})_{4} \otimes (\sigma_{z} \sigma_{x})_{5}$ \\
$\vert 1 \rangle_{1}\,\, \vert \psi^{-} \rangle_{23\:\:\:}$ & $\:\:\:\gamma
 \vert 00\rangle_{45} + \alpha \vert 11 \rangle_{45}\:\:\:$ &   apply 
$(\sigma_{x})_{4}\otimes (\sigma_{x})_{5}$ \\ 
\end{tabular}
\vspace{0.5cm}

The remaining seven possible quantum channels and seven sets of instructions
 are listed in Appendix~A. As it is shown in the same appendix, there are
eight possible ways also for teleporting the state $\beta \vert 10 \rangle + 
\delta \vert 01 \rangle$, but only four possible ways for teleporting
 factorizable (non entangled) states like  $\beta \vert 10 \rangle + \gamma
 \vert 11 \rangle $ and $\alpha \vert 00 \rangle + \delta \vert 01 \rangle$.

It is also possible to show that the machinery doesn't work on
 right-factorizable states like  $\alpha \vert 00 \rangle + \beta \vert 10
 \rangle $ and $\delta \vert 01 \rangle + \gamma \vert 11 \rangle $
 (see appendix~B).

\newpage 

Before concluding this section it is worthwhile summarizing the obtained
 results in a table$\,$:

\vspace{0,7cm}
\begin{tabular}{c|c}
Two-particle states $\:\:\:\:\:\:\:\:$ & $\:\:\:\:\;\:\:\:\:$
 Can be teleported? \\
\hline
& \\ 
$\alpha \vert 00 \rangle_{12} + \beta \vert 10 \rangle_{12} + \delta\vert 01
 \rangle_{12} + \gamma \vert 11 \rangle_{12}$ & No \\
$\alpha \vert 00 \rangle +\gamma \vert 11 \rangle $ & Yes, in eight different
 ways \\
$\beta \vert 10 \rangle +\delta \vert 01 \rangle $ & Yes, in eight different
 ways \\
$\beta \vert 10 \rangle +\gamma \vert 11 \rangle $ & Yes, in four different
 ways \\
$\alpha \vert 00 \rangle +\delta \vert 01 \rangle $ & Yes, in four different
 ways \\
$\alpha \vert 00 \rangle +\beta \vert 10 \rangle $ & No \\
$\delta \vert 01 \rangle +\gamma \vert 11 \rangle$ & No \\
\end{tabular}

\section{Conclusions}

In this work it has been shown that the most general and unknown two-particle
entangled state ( i.e. the state $\vert \psi \rangle = \alpha \vert 00 \rangle + 
\beta \vert 10 \rangle + \delta \vert 01 \rangle + \gamma \vert 11 \rangle$
with $\alpha, \beta, \gamma, \delta \neq 0\,$) cannot be teleported using 
only one (three-particle) channel and Bell measurements.

We have nevertheless shown that some two-particle entangled states, 
belonging to two dimensional subspaces of the whole Hilbert space, can be
successfully teleported from Alice to Bob using suitable and different 
three-particle quantum
channels, with the aim of Hadamard transformation, Bell measurements and
classical communication. 

We have listed which are the states and the sets of unitary transformations to
be performed by Bob in order to recreate a perfect copy of the original state,
without violating special relativity constraints ( classical communication
prevents in fact from sending faster-than-light messages ) and
no-cloning theorem ( the original state possessed by Alice is destroyed
 by Bell measurement).

%-------------------------------------------------------------------------

\section*{Appendix A}

Let us ask which are the states transmitted from Alice to Bob permitting him
 to reconstruct the original one.
The operations Bob may use are$\,$:

\begin{enumerate}
\item to do nothing, if the teleported state is already the original one; 
\item to make an exchange $\vert 0 \rangle \:\leftrightarrow \:\vert 1\rangle$
 on particle $4$ or on particle $5$, or on both;
\item to make a transformation of the $CNOT$ type; 
\item to use a product of $\vert 0 \rangle \:\leftrightarrow\:\vert 1\rangle$
 exchange and $CNOT$ transformation.
\end{enumerate}

 The unitary operator $CNOT$, which acts on two-particle states by reversing
 the second entry if the first is $1$, is defined as$\,$:

\begin{equation}
CNOT \,\vert 00 \rangle = \vert 00 \rangle \:\:\:\:\:\:\:\:\:\:  CNOT \,\vert 01 
\rangle = \vert 01 \rangle 
\end{equation}
\[   CNOT \,\vert 10 \rangle = \vert 11 \rangle \:\:\:\:\:\:\:\:\:\:  CNOT 
\,\vert 11 \rangle = \vert 10 \rangle  \]

Then, we can find out the states sent to Bob permitting teleportation by simply
 applying all the possible inverse operations to the original state.
 The possibilities, together with the operations which must be done by Bob,
 are listed below.

\vspace{0.7cm}
\begin{tabular}{cccc|c}
$ \vert 00 \rangle_{45}$ & $\vert 10 \rangle_{45}$ & $\vert 01 \rangle_{45}$
 & $\vert 11 \rangle_{45}$ &   Bob's instructions \\
\hline
$\alpha$ & $\beta$ & $\delta$ & $\gamma$ & do nothing \\
$\beta$ & $\alpha$ & $\gamma$ & $\delta$ & apply $(\sigma_{x})_{4}$ \\
$\delta$ & $\gamma$ & $\alpha$ & $\beta$ & apply $(\sigma_{x})_{5}$ \\
$\gamma$ & $\delta$ & $\beta$ & $\alpha$ & apply $(\sigma_{x})_{4}\otimes
(\sigma_{x})_{5}$ \\
$\alpha$ & $\gamma$ & $\delta$ & $\beta$ & apply $CNOT$ \\
$\beta$ & $\delta$ & $\gamma$ & $\alpha$ & $(\sigma_{x})_{4}\,CNOT$ \\
$\delta$ & $\beta$ & $\alpha$ & $\gamma$ & $(\sigma_{x})_{5}\,CNOT$ \\
$\gamma$ & $\alpha$ & $\beta$ & $\delta$ & apply $(\sigma_{x})_{4}\otimes
(\sigma_{x})_{5}\, CNOT $ 
\end{tabular}
\vspace{0.7cm}

It turns out that there are eight possible channels for transmitting states
 $\alpha \vert 00\rangle + \gamma \vert 11 \rangle$ and $\beta \vert 10\rangle
 + \delta \vert 01 \rangle$. In fact, putting $(\beta = 0,\: \delta = 0)$ or 
$( \alpha =0,\: \gamma = 0)$ respectively, we obtain eight different transmitted 
states. In the case of states $\alpha \vert 00\rangle + \delta \vert 01
 \rangle$ and $\beta \vert 10\rangle + \gamma \vert 11 \rangle$,
 the annihilation of coefficients $\beta, \gamma$ and $\alpha, \delta$ reduces
 to four the different states from which Bob can restore the original ones.

We are now going to list other seven possible quantum channels $\vert 
\phi\rangle_{345} $ and relative sets of instructions, with whom Alice and Bob 
can accomplish successful teleportation of the particular state
 $\alpha \vert 00 \rangle + \gamma \vert 11 \rangle$.

The various channels, being characterized by different choices of the
 coefficients $a,b...h$ to be inserted in equation (\ref{conadamo}), are
 indicated in the following schemes together with the results of Alice's
 measurements, the collapsed state of particles $4$ and $5$ and the unitary
 transformation which Bob must perform in order to complete teleportation
 process.

\vspace{0.3cm}
\[      { \bf Quantum\:\: Channels}  \] 

\vspace{0.3cm}
\begin{enumerate}
\item  Quantum Channel $\vert 010 \rangle_{345} + \vert 101 \rangle_{345}$
       $\:\:\:\:\:\:\:\:\:c=f=1,\:\:\: a=b=d=e=g=h=0$
       
\vspace{0.5cm}
\begin{tabular}{c|c|c}
%\:
Alice's measurements     &      Bob's states    &     Bob's instructions \\
\hline
$\vert 0 \rangle_{1}\,\, \vert \phi^{+} \rangle_{23}\:\:\:$ & $\:\:\: \alpha 
\vert 10 \rangle_{45} + \gamma \vert 01 \rangle_{45}\:\:\:$ &   apply 
$(\sigma_{x})_{4}\otimes I_{5}$ \\
$\vert 1 \rangle_{1}\,\, \vert \phi^{+} \rangle_{23}\:\:\:$ & $\:\:\:\alpha 
\vert 10 \rangle_{45} - \gamma \vert 01 \rangle_{45}\:\:\:$ &   apply 
$(\sigma_{x})_{4}\otimes (\sigma_{z})_{5}$   \\
$\vert 0 \rangle_{1}\,\, \vert \phi^{-} \rangle_{23}\:\:\:$ & $\:\:\:\alpha 
\vert 10 \rangle_{45} - \gamma \vert 01 \rangle_{45}\:\:\: $ &   apply 
$(\sigma_{x})_{4}\otimes (\sigma_{z})_{5}$ \\
$\vert 1 \rangle_{1}\,\, \vert \phi^{-} \rangle_{23}\:\:\:$ & $\:\:\:\alpha 
\vert 10 \rangle_{45} + \gamma \vert 01 \rangle_{45}\:\:\:$ &   apply 
$(\sigma_{x})_{4}\otimes I_{5}$  \\
$\vert 0 \rangle_{1}\,\, \vert \psi^{+} \rangle_{23}\:\:\:$ & $\:\:\:\gamma 
\vert 10 \rangle_{45} + \alpha \vert 01 \rangle_{45\:\:\:}$ &   apply 
$I_{4}\otimes (\sigma_{x})_{5}$  \\
$\vert 1 \rangle_{1}\,\, \vert \psi^{+} \rangle_{23}\:\:\:$ & $\:\:\:-\gamma 
\vert 10\rangle_{45} + \alpha \vert 01 \rangle_{45}\:\:\:$ &   apply 
$(\sigma_{z})_{4} \otimes
 (\sigma_{x})_{5} $ \\
$\vert 0 \rangle_{1}\,\, \vert \psi^{-} \rangle_{23}\:\:\:$ & $\:\:\:-\gamma 
\vert 10 \rangle_{45} + \alpha \vert 01 \rangle_{45}\:\:\:$ &   apply 
$(\sigma_{z})_{4} \otimes (\sigma_{x})_{5} $ \\
$\vert 1 \rangle_{1}\,\, \vert \psi^{-} \rangle_{23\:\:\:}$ & $\:\:\:\gamma
 \vert 10 \rangle_{45} + \alpha \vert 01 \rangle_{45}\:\:\:$ &   apply $I_{4}
\otimes (\sigma_{x})_{5} $ 
\end{tabular}
\newpage

\item  Quantum Channel $\vert 100 \rangle_{345} + \vert 011 \rangle_{345}$
       $\:\:\:\:\:\:\:\:\:b=g=1,\:\:\: a=c=d=e=f=h=0$
       
\vspace{0.2cm}
\begin{tabular}{c|c|c}
Alice's measurements     &      Bob's states    &     Bob's instructions \\
\hline
$| 0 \rangle_{1}\,\, \vert \phi^{+} \rangle_{23}\:\:\:$ & $\:\:\:\gamma \vert 00 
\rangle_{45} + \alpha \vert 11 \rangle_{45\:\:\:}$ &   apply $(\sigma_{x})_{4} 
\otimes
 (\sigma_{x})_{5} $   \\
$\vert 1 \rangle_{1}\,\, \vert \phi^{+} \rangle_{23}\:\:\:$ & $\:\:\:-\gamma 
\vert 00\rangle_{45} + \alpha \vert 11 \rangle_{45}\:\:\:$ &   apply 
$(\sigma_{z} \sigma_{x})_{4} \otimes (\sigma_{x})_{5} $  \\
$\vert 0 \rangle_{1}\,\, \vert \phi^{-} \rangle_{23}\:\:\:$ & $\:\:\:-\gamma 
\vert 00 \rangle_{45} + \alpha \vert 11 \rangle_{45}\:\:\:$ &   apply 
$(\sigma_{z} \sigma_{x})_{4} \otimes (\sigma_{x})_{5} $   \\
$\vert 1 \rangle_{1}\,\, \vert \phi^{-} \rangle_{23\:\:\:}$ & $\:\:\:\gamma
 \vert 00 \rangle_{45} + \alpha \vert 11 \rangle_{45}\:\:\:$ &   apply 
$(\sigma_{x})_{4} \otimes (\sigma_{x})_{5} $    \\
$\vert 0 \rangle_{1}\,\, \vert \psi^{+} \rangle_{23}\:\:\:$ & $\:\:\: \alpha 
\vert 00 \rangle_{45} + \gamma \vert 11 \rangle_{45}\:\:\:$ &   do nothing \\
$\vert 1 \rangle_{1}\,\, \vert \psi^{+} \rangle_{23}\:\:\:$ & $\:\:\:\alpha 
\vert 00 \rangle_{45} - \gamma \vert 11 \rangle_{45}\:\:\:$ &   apply 
$I_{4}\otimes (\sigma_{z})_{5}$ \\
$\vert 0 \rangle_{1}\,\, \vert \psi^{-} \rangle_{23}\:\:\:$ & $\:\:\:\alpha 
\vert 00 \rangle_{45} - \gamma \vert 11 \rangle_{45}\:\:\: $ &   apply 
$I_{4}\otimes (\sigma_{z})_{5}$ \\
$\vert 1 \rangle_{1}\,\, \vert \psi^{-} \rangle_{23}\:\:\:$ & $\:\:\:\alpha 
\vert 00 \rangle_{45} + \gamma \vert 11 \rangle_{45}\:\:\:$ &    do nothing   
\end{tabular}
\vspace{0.3cm}
\item  Quantum Channel $\vert 110 \rangle_{345} + \vert 001 \rangle_{345}$
       $\:\:\:\:\:\:\:\:\:d=e=1,\:\:\:a=b=c=f=g=h=0$ 
       
\vspace{0.2cm}
\begin{tabular}{c|c|c}
Alice's measurements     &      Bob's states    &     Bob's instructions \\
\hline
$\vert 0 \rangle_{1}\,\, \vert \phi^{+} \rangle_{23}\:\:\:$ & $\:\:\:\gamma 
\vert 10 \rangle_{45} + \alpha \vert 01 \rangle_{45\:\:\:}$ &   apply 
$I_{4}\otimes (\sigma_{x})_{5}$   \\
$\vert 1 \rangle_{1}\,\, \vert \phi^{+} \rangle_{23}\:\:\:$ & $\:\:\:-\gamma 
\vert 10\rangle_{45} + \alpha \vert 01 \rangle_{45}\:\:\:$ &   apply 
$(\sigma_{z})_{4} \otimes (\sigma_{x})_{5} $   \\
$\vert 0 \rangle_{1}\,\, \vert \phi^{-} \rangle_{23}\:\:\:$ & $\:\:\:-\gamma 
\vert 10 \rangle_{45} + \alpha \vert 01 \rangle_{45}\:\:\:$ &   apply 
$(\sigma_{z})_{4} \otimes (\sigma_{x})_{5} $ \\
$\vert 1 \rangle_{1}\,\, \vert \phi^{-} \rangle_{23\:\:\:}$ & $\:\:\:\gamma
 \vert 10 \rangle_{45} + \alpha \vert 01 \rangle_{45}\:\:\:$ &   apply 
$I_{4}\otimes (\sigma_{x})_{5}$      \\
$\vert 0 \rangle_{1}\,\, \vert \psi^{+} \rangle_{23}\:\:\:$ & $\:\:\: \alpha 
\vert 10 \rangle_{45} + \gamma \vert 01 \rangle_{45}\:\:\:$ &   apply 
$(\sigma_{x})_{4}\otimes I_{5}$ \\
$\vert 1 \rangle_{1}\,\, \vert \psi^{+} \rangle_{23}\:\:\:$ & $\:\:\:\alpha 
\vert 10 \rangle_{45} - \gamma \vert 01 \rangle_{45}\:\:\:$ &   apply 
$(\sigma_{x})_{4} \otimes (\sigma_{z})_{5} $   \\
$\vert 0 \rangle_{1}\,\, \vert \psi^{-} \rangle_{23}\:\:\:$ & $\:\:\:\alpha 
\vert 10 \rangle_{45} - \gamma \vert 01 \rangle_{45}\:\:\: $ &   apply 
$(\sigma_{x})_{4} \otimes (\sigma_{z})_{5} $   \\
$\vert 1 \rangle_{1}\,\, \vert \psi^{-} \rangle_{23}\:\:\:$ & $\:\:\:\alpha 
\vert 10 \rangle_{45} + \gamma \vert 01 \rangle_{45}\:\:\:$ &    apply 
$(\sigma_{x})_{4}\otimes I_{5}$   \\ 
\end{tabular}
\vspace{0.3cm}

\item  Quantum Channel $\vert 000 \rangle_{345} + \vert 110 \rangle_{345}$
       $\:\:\:\:\:\:\:\:\:a=e=1,\:\:\: b=c=d=f=g=h=0$

\vspace{0.2cm}
\begin{tabular}{c|c|c}
Alice's measurements     &      Bob's states    &     Bob's instructions \\
\hline
$\vert 0 \rangle_{1}\,\, \vert \phi^{+} \rangle_{23}\:\:\:$ & $\:\:\: \alpha 
\vert 00 \rangle_{45} + \gamma \vert 10 \rangle_{45}\:\:\:$ &   apply $CNOT$ \\
$\vert 1 \rangle_{1}\,\, \vert \phi^{+} \rangle_{23}\:\:\:$ & $\:\:\:\alpha 
\vert 00 \rangle_{45} - \gamma \vert 10 \rangle_{45}\:\:\:$ &   apply 
$(\sigma_{z})_{5}\,CNOT$   \\
$\vert 0 \rangle_{1}\,\, \vert \phi^{-} \rangle_{23}\:\:\:$ & $\:\:\:\alpha 
\vert 00
 \rangle_{45} - \gamma \vert 10 \rangle_{45}\:\:\: $ &   apply 
$(\sigma_{z})_{5}\,CNOT$    \\
$\vert 0 \rangle_{1}\,\, \vert \phi^{-} \rangle_{23}\:\:\:$ & $\:\:\:\alpha 
\vert 00 \rangle_{45} + \gamma \vert 10 \rangle_{45}\:\:\: $ &   apply $CNOT$ \\
$\vert 0 \rangle_{1}\,\, \vert \psi^{+} \rangle_{23}\:\:\:$ & $\:\:\:\gamma 
\vert 00 \rangle_{45} + \alpha \vert 10 \rangle_{45\:\:\:}$ &   apply $ 
(\sigma_{x})_{4} \otimes
 (\sigma_{x})_{5}\,CNOT $  \\
$\vert 1 \rangle_{1}\,\, \vert \psi^{+} \rangle_{23}\:\:\:$ & $\:\:\:-\gamma 
\vert 00\rangle_{45} + \alpha \vert 10 \rangle_{45}\:\:\:$ &   apply 
$(\sigma_{z}\sigma_{x})_{4} \otimes (\sigma_{x})_{5}\,CNOT $ \\
$\vert 0 \rangle_{1}\,\, \vert \psi^{-} \rangle_{23}\:\:\:$ & $\:\:\:-\gamma 
\vert 00 \rangle_{45} + \alpha \vert 10 \rangle_{45}\:\:\:$ &   apply  
$(\sigma_{z}\sigma_{x})_{4} \otimes (\sigma_{x})_{5}\,CNOT $ \\
$\vert 1 \rangle_{1}\,\, \vert \psi^{-} \rangle_{23\:\:\:}$ & $\:\:\:\gamma 
\vert 00
 \rangle_{45} + \alpha \vert 10 \rangle_{45}\:\:\:$ &   apply $  
(\sigma_{x})_{4} \otimes (\sigma_{x})_{5}\,CNOT $  
\end{tabular}
\vspace{0.3cm}

\item  Quantum Channel $\vert 100 \rangle_{345} + \vert 010 \rangle_{345}$
       $\:\:\:\:\:\:\:\:\:b=c=1, \:\:\: a=d=e=f=g=h=0$

\vspace{0.2cm}
\begin{tabular}{c|c|c}
Alice's measurements     &      Bob's states    &     Bob's instructions \\
\hline
$\vert 0 \rangle_{1}\,\, \vert \phi^{+} \rangle_{23}\:\:\:$ & $\:\:\:\gamma 
\vert 00 \rangle_{45} + \alpha \vert 10 \rangle_{45\:\:\:}$ &   apply $ 
(\sigma_{x})_{4} \otimes
 (\sigma_{x})_{5}\,CNOT $  \\
$\vert 1 \rangle_{1}\,\, \vert \phi^{+} \rangle_{23}\:\:\:$ & $\:\:\:-\gamma 
\vert 00\rangle_{45} + \alpha \vert 10 \rangle_{45}\:\:\:$ &   apply 
$(\sigma_{z}\sigma_{x})_{4} \otimes (\sigma_{x})_{5}\,CNOT $ \\
$\vert 0 \rangle_{1}\,\, \vert \phi^{-} \rangle_{23}\:\:\:$ & $\:\:\:-\gamma 
\vert 00 \rangle_{45} + \alpha \vert 10 \rangle_{45}\:\:\:$ &   apply  
$(\sigma_{z}\sigma_{x})_{4} \otimes (\sigma_{x})_{5}\,CNOT $ \\
$\vert 1 \rangle_{1}\,\, \vert \phi^{-} \rangle_{23\:\:\:}$ & $\:\:\:\gamma 
\vert 00
 \rangle_{45} + \alpha \vert 10 \rangle_{45}\:\:\:$ &   apply $  
(\sigma_{x})_{4} \otimes (\sigma_{x})_{5}\,CNOT $  \\
$\vert 0 \rangle_{1}\,\, \vert \psi^{+} \rangle_{23}\:\:\:$ & $\:\:\: \alpha 
\vert 00 \rangle_{45} + \gamma \vert 10 \rangle_{45}\:\:\:$ &   apply $CNOT$ \\
$\vert 1 \rangle_{1}\,\, \vert \psi^{+} \rangle_{23}\:\:\:$ & $\:\:\:\alpha 
\vert 00 \rangle_{45} - \gamma \vert 10 \rangle_{45}\:\:\:$ &   apply 
$(\sigma_{z})_{5}\,CNOT$   \\
$\vert 0 \rangle_{1}\,\, \vert \psi^{-} \rangle_{23}\:\:\:$ & $\:\:\:\alpha 
\vert 00
 \rangle_{45} - \gamma \vert 10 \rangle_{45}\:\:\: $ &   apply $(\sigma_{z})_{5}
\,CNOT$    \\
$\vert 0 \rangle_{1}\,\, \vert \psi^{-} \rangle_{23}\:\:\:$ & $\:\:\:\alpha 
\vert 00
 \rangle_{45} + \gamma \vert 10 \rangle_{45}\:\:\: $ &   apply $CNOT$ \\
\end{tabular}

\newpage
\item  Quantum Channel $\vert 101 \rangle_{345} + \vert 011 \rangle_{345}$
       $\:\:\:\:\:\:\:\:\:f=g=1, \:\:\: a=b=c=d=e=h=0$

\vspace{0.2cm}
\begin{tabular}{c|c|c}
Alice's measurements     &      Bob's states    &     Bob's instructions \\
\hline
$\vert 0 \rangle_{1}\,\, \vert \phi^{+} \rangle_{23}\:\:\:$ & $\:\:\:\gamma 
\vert 01 \rangle_{45} + \alpha \vert 11 \rangle_{45}\:\:\:$ &   apply $ 
(\sigma_{x})_{4}\, CNOT $  \\
$\vert 1 \rangle_{1}\,\, \vert \phi^{+} \rangle_{23}\:\:\:$ & $\:\:\:-\gamma 
\vert 01\rangle_{45} + \alpha \vert 11 \rangle_{45}\:\:\:$ &   apply $ 
(\sigma_{z})_{5} \otimes (\sigma_{x})_{4}\,CNOT $ \\
$\vert 0 \rangle_{1}\,\, \vert \phi^{-} \rangle_{23}\:\:\:$ & $\:\:\:-\gamma 
\vert 01 \rangle_{45} + \alpha \vert 11 \rangle_{45}\:\:\:$ &   apply $ 
(\sigma_{z})_{5} \otimes (\sigma_{x})_{4}\,CNOT $ \\  
$\vert 1 \rangle_{1}\,\, \vert \phi^{-} \rangle_{23\:\:\:}$ & $\:\:\:\gamma
 \vert 01 \rangle_{45} + \alpha \vert 11 \rangle_{45}\:\:\:$ &   apply 
$(\sigma_{x})_{4}\,CNOT $  \\
$\vert 0 \rangle_{1}\,\, \vert \psi^{+} \rangle_{23}\:\:\:$ & $\:\:\: \alpha 
\vert 01 \rangle_{45} + \gamma \vert 11 \rangle_{45}\:\:\:$ &   apply 
$(\sigma_{x})_{5}\,CNOT $  \\
$\vert 1 \rangle_{1}\,\, \vert \psi^{+} \rangle_{23}\:\:\:$ & $\:\:\:\alpha 
\vert 01 \rangle_{45} - \gamma \vert 11 \rangle_{45}\:\:\:$ &   apply $ 
(\sigma_{z})_{4} \otimes (\sigma_{x})_{5}\,CNOT $ \\
$\vert 0 \rangle_{1}\,\, \vert \psi^{-} \rangle_{23}\:\:\:$ & $\:\:\:\alpha 
\vert 01 \rangle_{45} - \gamma \vert 11 \rangle_{45}\:\:\: $ &   apply $ 
(\sigma_{z})_{4} \otimes (\sigma_{x})_{5}\,CNOT $ \\
$\vert 0 \rangle_{1}\,\, \vert \psi^{-} \rangle_{23}\:\:\:$ & $\:\:\:\alpha 
\vert 01 \rangle_{45} + \gamma \vert 11 \rangle_{45}\:\:\: $ &   apply $ 
(\sigma_{x})_{5}\, CNOT $  \\
\end{tabular}
\vspace{0.3cm}
\item  Quantum Channel $\vert 001 \rangle_{345} + \vert 111 \rangle_{345}$
       $\:\:\:\:\:\:\:\:\:d=h=1, \:\:\: a=b=c=e=f=g=0$

\vspace{0.2cm}
\begin{tabular}{c|c|c}
Alice's measurements     &      Bob's states    &     Bob's instructions \\
\hline
$\vert 0 \rangle_{1}\,\, \vert \phi^{+} \rangle_{23}\:\:\:$ & $\:\:\: \alpha 
\vert 01 \rangle_{45} + \gamma \vert 11 \rangle_{45}\:\:\:$ &   apply 
$(\sigma_{x})_{5}\,CNOT $  \\
$\vert 1 \rangle_{1}\,\, \vert \phi^{+} \rangle_{23}\:\:\:$ & $\:\:\:\alpha 
\vert 01 \rangle_{45} - \gamma \vert 11 \rangle_{45}\:\:\:$ &   apply $ 
(\sigma_{z})_{4} \otimes (\sigma_{x})_{5}\,CNOT $ \\
$\vert 0 \rangle_{1}\,\, \vert \phi^{-} \rangle_{23}\:\:\:$ & $\:\:\:\alpha 
\vert 01 \rangle_{45} - \gamma \vert 11 \rangle_{45}\:\:\: $ &   apply $ 
(\sigma_{z})_{4} \otimes (\sigma_{x})_{5}\,CNOT $ \\
$\vert 0 \rangle_{1}\,\, \vert \phi^{-} \rangle_{23}\:\:\:$ & $\:\:\:\alpha 
\vert 01 \rangle_{45} + \gamma \vert 11 \rangle_{45}\:\:\: $ &   apply $ 
(\sigma_{x})_{5}\, CNOT $  \\
$\vert 0 \rangle_{1}\,\, \vert \psi^{+} \rangle_{23}\:\:\:$ & $\:\:\:\gamma 
\vert 01 \rangle_{45} + \alpha \vert 11 \rangle_{45}\:\:\:$ &   apply $ 
(\sigma_{x})_{4}\, CNOT $  \\
$\vert 1 \rangle_{1}\,\, \vert \psi^{+} \rangle_{23}\:\:\:$ & $\:\:\:-\gamma 
\vert 01\rangle_{45} + \alpha \vert 11 \rangle_{45}\:\:\:$ &   apply $ 
(\sigma_{z})_{5} \otimes (\sigma_{x})_{4}\,CNOT $ \\
$\vert 0 \rangle_{1}\,\, \vert \psi^{-} \rangle_{23}\:\:\:$ & $\:\:\:-\gamma 
\vert 01 \rangle_{45} + \alpha \vert 11 \rangle_{45}\:\:\:$ &   apply $ 
(\sigma_{z})_{5} \otimes (\sigma_{x})_{4}\,CNOT $ \\  
$\vert 1 \rangle_{1}\,\, \vert \psi^{-} \rangle_{23\:\:\:}$ & $\:\:\:\gamma
 \vert 01 \rangle_{45} + \alpha \vert 11 \rangle_{45}\:\:\:$ &   apply 
$(\sigma_{x})_{4}\,CNOT $  \\
\end{tabular}
\end{enumerate}

\section*{Appendix B}

 For the sake of simplicity we will only enumerate the permitted three-particle
quantum channels that Alice and Bob may use to teleport the following
two-particle states$\,$:

\vspace{0.5cm}
\begin{tabular}{c|c}
Two-particle states $\:\:\:\:\:\:\:$ & $\:\:\:\:\:\:\:$  Quantum channels  \\
\hline
& \\
$[\,\beta \vert 10 \rangle + \delta \vert 01 \rangle]_{12} $  & $\:\:\:\:\:\:\:$ 
$( \vert 010 \rangle + \vert 101\rangle )_{345}$ \\
 &$\:\:\:\:\:\:\:$ $( \vert 000 \rangle  + \vert 111\rangle )_{345}$ \\
 &$\:\:\:\:\:\:\:$ $( \vert 100 \rangle  + \vert 011\rangle )_{345}$ \\
 &$\:\:\:\:\:\:\:$ $( \vert 110 \rangle  + \vert 001\rangle )_{345}$ \\
 &$\:\:\:\:\:\:\:$ $( \vert 001 \rangle  + \vert 111\rangle )_{345}$ \\
 &$\:\:\:\:\:\:\:$ $( \vert 101 \rangle  + \vert 011\rangle )_{345}$ \\
 &$\:\:\:\:\:\:\:$ $( \vert 000 \rangle  + \vert 110\rangle )_{345}$ \\
 &$\:\:\:\:\:\:\:$ $( \vert 100 \rangle  + \vert 010\rangle )_{345}$ \\
& \\
$[\,\alpha \vert 00 \rangle + \delta \vert 01 \rangle]_{12} $  &$\:\:\:\:\:\:\:$ 
 $( \vert 000 \rangle + \vert 110\rangle )_{345}$ \\
&$\:\:\:\:\:\:\:$ $( \vert 100 \rangle  + \vert 010\rangle )_{345}$ \\
&$\:\:\:\:\:\:\:$ $( \vert 001 \rangle  + \vert 111\rangle )_{345}$ \\
&$\:\:\:\:\:\:\:$ $( \vert 101 \rangle  + \vert 011\rangle )_{345}$ \\
\end{tabular}

\begin{tabular}{c|c}
Two-particle states $\:\:\:\:\:\:\:$ & $\:\:\:\:\:\:\:$  Quantum channels  \\
\hline
& \\
$[\,\beta \vert 10 \rangle + \gamma \vert 11 \rangle]_{12}$ & $\:\:\:\:\:\:\:$
  $( \vert 000 \rangle + \vert 110\rangle )_{345}$ \\
&$\:\:\:\:\:\:\:$ $( \vert 100 \rangle  + \vert 010\rangle )_{345}$ \\
&$\:\:\:\:\:\:\:$ $( \vert 001 \rangle  + \vert 111\rangle )_{345}$ \\
&$\:\:\:\:\:\:\:$ $( \vert 101 \rangle  + \vert 011\rangle )_{345}$ 
 \end{tabular}

\vspace{0.5cm}

Two-particle factorized states in which the unknown part is in channel $1$ ( 
i.e., states $\alpha \vert 00 \rangle_{12} + \beta \vert 10 \rangle_{12} \equiv 
( 
\alpha \vert 0 \rangle_{1} + \beta \vert 1\rangle_{1} )\, \vert 0 \rangle_{2}$ 
and  
$\delta \vert  01 \rangle_{12} + \gamma \vert 11 \rangle_{12} \equiv ( \delta 
\vert 0 \rangle_{1} + \gamma \vert 1\rangle_{1} )\, \vert 1\rangle_{2}$ cannot 
be transmitted using the present method.

 Let us consider for example the state $\alpha \vert 00 \rangle_{12} + \beta
 \vert 10 \rangle_{12}\:$; this means to choose $\gamma = 0$ and $\delta = 0$
 in equation (\ref{conadamo}), which becomes$\,$:

\begin{eqnarray}
\vert \tilde{\Omega} \rangle & = & \frac{1}{2\sqrt{N}}\, [ \,(\alpha + \beta) 
\vert 0
 \rangle_{1} + ( \alpha - \beta) \vert 1 \rangle_{1}\, ]\,\cdot [\,  \vert 
\phi^{+}
 \rangle_{23} +\vert \phi^{-} \rangle_{23} + \vert \psi^{+} \rangle_{23} +
 \vert \psi^{-}  \rangle_{23}\, ] \cdot \nonumber \\
 & & \:\:\:\:\:\:\:\:\:\:\:\: [\, a \vert 00 \rangle_{45} + c \vert 10 
\rangle_{45} + d \vert 01
 \rangle_{45} + g \vert 11 \rangle_{45}\, ]
\end{eqnarray}

and the teleporting method cannot be applied.
The unknown state must be put in channel $2$, which is involved in the Bell 
measurement. However, as already noted, in such a case one can resort to the
 original standard teleportation procedure.


\begin{thebibliography}{99}

\bibitem{ref1} C.H.Bennett, G.Brassard, C.Crepeau, R.Josza, A.Peres and 
W.K.Wootters, {\it Phys. Rev. Lett.} {\bf 70}, 1895 (1993).

\bibitem{ref2} W.K.Wootters, W.H.Zurek, {\it Nature}, {\bf 299}, 802 (1982);\\
G.C.Ghirardi, T.Weber, {\it Il Nuovo Cimento}, {\bf 78B}, 9 (1983).

\bibitem{ref3} D.Bouwmeester, J.Pan, M.Daniell, H.Weinfurter, A.Zeilinger, 
preprint;
R.J.Nelson, D.G.Cory, S.Lloyd, preprint.

\end{thebibliography}
\end{document}